\documentclass[12pt]{article}
\newcommand{\bq}{\begin{equation}}
\newcommand{\eq}{\end{equation}}
\newcommand{\ba}{\begin{eqnarray}}
\newcommand{\ea}{\end{eqnarray}}

\newcommand{\nl }{ \nonumber  }

\newcommand{\p}{\partial}
\newcommand{\h}{\hspace{.5cm}}
\newcommand{\s}{\sigma}

\def\appendix#1{
  \setcounter{equation}{0}
  \renewcommand{\thesection}{\Alph{section}}
  \section*{Appendix \thesection\protect\indent \parbox[t]{11.15cm}
  {#1} }
  \addcontentsline{toc}{section}{Appendix \thesection\ \ \ #1}
  }

\begin{document}
\vspace*{.5cm}
\begin{center}
{\bf ON THE CLASSICAL STRING SOLUTIONS AND STRING/FIELD THEORY
DUALITY II \vspace*{0.5cm}\\ D. Aleksandrova, P. Bozhilov}
\\ {\it Department of Theoretical and Applied Physics, \\
Shoumen University, 9712 Shoumen, Bulgaria\\
E-mail:} p.bozhilov@shu-bg.net
\end{center}
\vspace*{0.5cm}

Based on the recently considered classical string configurations,
in the framework of the semi-classical limit of the string/gauge
theory correspondence, we describe a procedure for obtaining exact
classical string solutions in general string theory backgrounds,
when the string embedding coordinates depend non-linearly on the
worldsheet spatial parameter. The tensionless limit, corresponding
to small t'Hooft coupling on the field theory side, is also
considered. Applying the developed approach, we first reproduce
some known results. Then, we find new string solutions - with two
spins in $AdS_5$ black hole background and in $AdS_5\times S^5$
with two spins and up to nine independent conserved $R$-charges.

\vspace*{.5cm} {\bf PACS codes:} 11.25.-w, 11.27.+d, 11.30.-j

\vspace*{.5cm} {\bf Keywords:} Bosonic Strings, AdS-CFT and dS-CFT
Correspondence, Space-Time Symmetries, Integrable Equations in
Physics.

\vspace*{.5cm}

\section{Introduction}
The article \cite{4} on the semi-classical limit of the
string/gauge theory duality initiated an interest in the
investigation of the connection between the classical string
solutions, their semi-classical quantization and the string/field
theory correspondence \cite{5} - \cite{34} \footnote{For earlier
studies on the subject, see \cite{NSetall,PB01} and the references in
\cite{PB01} and \cite{NS03}.}. Most of the papers consider
different string configurations in type IIB $AdS_5\times S^5$
background. However, the string dynamics has been investigated in
other string theory backgrounds, known to have field theory dual
descriptions in different dimensions, with different number of (or
without) supersymmetries, conformal or non-conformal.

For establishing the correspondence between the semi-classically
quantized string solutions and the appropriate objects in the dual
field theory, it is essential for one to know the explicit
expressions for the conserved quantities like energy, angular
momentum, etc., on the string theory side. Their existence is
connected with the symmetries of the corresponding supergravity
backgrounds. The analysis of the connection between the ansatzes,
used to obtain exact string solutions, and the background
symmetries shows that the string embeddings can be divided into the
following four types \cite{AB}:
\ba\label{tLA} X^\mu (\tau, \sigma)= \Lambda^\mu_0\tau
+ \Lambda^\mu_1\sigma,\h X^a (\tau, \sigma)= Y^a (\tau);\\
\label{tGA} X^\mu (\tau, \sigma)= \Lambda^\mu_0\tau +
\Lambda^\mu_1\sigma + Y^\mu (\tau),\h X^a (\tau, \sigma)= Y^a
(\tau);\\ \label{sLA} X^\mu (\tau, \sigma)= \Lambda^\mu_0\tau +
\Lambda^\mu_1\sigma,\h X^a (\tau, \sigma)= Z^a (\sigma);\\
\label{sGA} X^\mu (\tau, \sigma)= \Lambda^\mu_0\tau +
\Lambda^\mu_1\sigma + Z^\mu (\sigma),\h X^a (\tau, \sigma)= Z^a
(\sigma);\\ \nl \Lambda^\mu_m = const,\h (m=0,1).\ea Here, the
string embedding coordinates $X^M(\tau,\sigma)$, $(M=0,1,\ldots,
D-1)$, are represented as $X^M=(X^\mu,X^a)$. $X^\mu(\tau,\sigma)$
correspond to the space-time coordinates $x^\mu$, on which the
background fields do not depend. This means that there exist
$n_\mu$ commuting Killing vectors $\p/\p x^\mu$, where $n_\mu$ is
the number of the coordinates $x^\mu$.

All the ansatzes used in \cite{4} - \cite{34} for the string
embedding, are particular cases of (\ref{tLA}) - (\ref{sGA}),
except those in \cite{21,34}\footnote{See section 5.}. In
\cite{11,14,19,28} there are of type (\ref{tLA}), in \cite{4} and
\cite{12} - of type (\ref{tGA}). In \cite{4} - \cite{9},
\cite{12}, \cite{13}, \cite{27}, \cite{15}, \cite{19} - \cite{25},
\cite{29} - \cite{33}, ansatzes of type (\ref{sLA}) are
considered. Solutions, based on the ansatzes of type (\ref{sGA}),
are obtained in \cite{16} - \cite{19}, \cite{26}.

The aim of this article is to describe a procedure for obtaining
the exact classical string solutions
in general string theory backgrounds, based on the ansatzes
(\ref{sLA}) and (\ref{sGA})\footnote{The corresponding results,
based on the ansatzes (\ref{tLA}) and (\ref{tGA}), can be found in
\cite{AB}.}. We will use more general worldsheet gauge than the
conformal one, in order to be able to discuss the tensionless
limit $T\to 0$, corresponding to small t'Hooft coupling
$\lambda\to 0$ on the field theory side.

Let us also note that in (\ref{tLA})-(\ref{sGA}), we have
separated the cases $Y^\mu =0$ and $Y^\mu\ne 0$, $Z^\mu =0$ and
$Z^\mu\ne 0$, because the types of the string solutions in these
cases are essentially different, as we will see later on.

The paper is organized as follows. In Sec.2 and Sec.3, we describe
the string dynamics and give the corresponding exact solutions of
the equations of motion and constraints, based on the ansatzes
(\ref{sLA}) and (\ref{sGA}) respectively. Sec.4 is devoted to some
applications of the derived general results. In Sec. 5, we
conclude with several remarks.

\setcounter{equation}{0}
\section{Exact string solutions}
In our further considerations, we will explore the Polyakov type
action for the bosonic string in a $D$-dimensional curved
space-time with metric tensor $g_{MN}(x)$, interacting with a
background 2-form gauge field $b_{MN}(x)$ via Wess-Zumino term
\ba\label{pa} &&S^{P}=\int d^{2}\xi\mathcal{L}^P,\h \mathcal{L}^P
=-\frac{1}{2}\left(T\sqrt{-\gamma}\gamma^{mn}G_{mn}-Q\varepsilon^{mn}
B_{mn}\right),\\ \nl && \xi^m=(\xi^0,\xi^1)=(\tau,\s),\h m,n =
(0,1),\ea where  \ba\nl &&G_{mn}= \p_m X^M\p_n X^N g_{MN},\h
B_{mn}=\p_{m}X^{M}\p_{n}X^{N} b_{MN}, \\ \nl &&(\p_m=\p/\p\xi^m,\h
M,N = 0,1,\ldots,D-1),\ea are the fields induced on the string
worldsheet, $\gamma$ is the determinant of the auxiliary
worldsheet metric $\gamma_{mn}$, and $\gamma^{mn}$ is its inverse.
The position of the string in the background space-time is given
by $x^M=X^M(\xi^m)$, and $T=1/2\pi\alpha'$, $Q$ are the string
tension and charge, respectively. If we consider the action
(\ref{pa}) as a bosonic part of a supersymmetric one, we have to
put $Q=\pm T$. In what follows, $Q =T$.

The equations of motion for $X^M$ following from (\ref{pa}) are:
\ba \label{em}
&&-g_{LK}\left[\p_m\left(\sqrt{-\gamma}\gamma^{mn}\p_nX^K\right) +
\sqrt{-\gamma}\gamma^{mn}\Gamma^K_{MN}\p_m X^M\p_n X^N\right]\\
\nl &&=\frac{1}{2}H_{LMN}\epsilon^{mn}\p_m X^M\p_n X^N,\ea where
\ba\nl
&&\Gamma_{L,MN}=g_{LK}\Gamma^K_{MN}=\frac{1}{2}\left(\p_Mg_{NL}
+\p_Ng_{ML}-\p_Lg_{MN}\right),\\ \nl &&H_{LMN}= \p_L b_{MN}+ \p_M
b_{NL} + \p_N b_{LM},\ea are the components of the symmetric
connection corresponding to the metric $g_{MN}$, and the field
strength of the gauge field $b_{MN}$ respectively. The constraints
are obtained by varying the action (\ref{pa}) with respect to
$\gamma_{mn}$: \ba\label{oc} \delta_{\gamma_{mn}}S^P=0\Rightarrow
\left(\gamma^{kl}\gamma^{mn}-2\gamma^{km}\gamma^{ln}\right)G_{mn}=0.\ea

Now, our task is to find exact solutions of the nonlinear
differential equations (\ref{em}) and (\ref{oc}). Let us first
consider the constraints (\ref{oc}). We have three constraints in
(\ref{oc}), but only two of them are independent. To extract the
independent ones, we rewrite the three constraints as follows:
\ba\label{00}
&&\left(\gamma^{00}\gamma^{mn}-2\gamma^{0m}\gamma^{0n}\right)G_{mn}=0,\\
\label{01}
&&\left(\gamma^{01}\gamma^{mn}-2\gamma^{0m}\gamma^{1n}\right)G_{mn}=0,\\
\label{11}
&&\left(\gamma^{11}\gamma^{mn}-2\gamma^{1m}\gamma^{1n}\right)G_{mn}=0.\ea
Inserting $G_{11}$ from (\ref{11}) into (\ref{00}) and (\ref{01}),
one obtains that both of them are satisfied, when the equality
\ba\label{1} \gamma^{01}G_{00}+\gamma^{11}G_{01}=0 \ea holds. To
simplify the constraint (\ref{11}), we put (\ref{1}) in it, which
results in \ba\label{0} \gamma^{00}G_{00}-\gamma^{11}G_{11}=0.\ea
Thus, our {\it independent} constraints, with which we will work
from now on, are given by (\ref{1}) and (\ref{0}).

Now let us turn to the equations of motion (\ref{em}). We will use
the gauge $\gamma^{mn}=constants$, in which they simplify to
\ba\label{sem} g_{LK}\gamma^{mn} \left(\p_m\p_nX^K
+\Gamma^K_{MN}\p_m X^M\p_n X^N\right)=
-\frac{1}{2\sqrt{-\gamma}}H_{LMN}\epsilon^{mn}\p_m X^M \p_n X^N
.\ea In particular, $\gamma^{mn}=\eta^{mn}=diag(-1,1)$ correspond
to the commonly used {\it conformal gauge}.

\subsection{Solving the equations of motion and constraints}
In solving the equations of motion and constraints, we will use
string embedding that exploits the symmetries of the background.
Namely, our ansatz for the string coordinates
$X^M(\tau,\sigma)=(X^\mu, X^a)$ in this section is given by
(\ref{sLA}), and $x^\mu$ are the target space-time coordinates, on
which the background fields do not depend: \ba\label{ob}\p_\mu
g_{MN} =0,\h \p_\mu b_{MN} =0.\ea

Taking into account the ansatz (\ref{sLA}) and under the
conditions (\ref{ob}), one obtains the following reduced
Lagrangian density, arising from the action (\ref{pa}) (the prime
is used for $d/d\sigma$)\ba\label{LRa}\mathcal{L}^{A}(\sigma) =
-\frac{T}{2}\sqrt{- \gamma}\left[ \gamma^{11}g_{ab}Z'^aZ'^b+
2\left( \gamma^ {1m}\Lambda_m^\mu g_{\mu a} - \frac{1}{\sqrt{-
\gamma}}\Lambda_0^\mu b_{\mu a} \right)Z'^a \right.+ \\ \nl \left.
+ \gamma^{mn}\Lambda_m^\mu \Lambda_n^\nu g_{\mu \nu}-
\frac{2}{\sqrt{-\gamma}}\Lambda_0^\mu \Lambda_1^\nu b_{\mu \nu}
\right],\ea where the fields induced on the string worldsheet are
given by \ba\nl &&G_{00}=\Lambda^\mu_0\Lambda^\nu_0 g_{\mu\nu},\h
G_{01}=\Lambda^\mu_0\left(g_{\mu a}Z'^a + \Lambda^\nu_1
g_{\mu\nu}\right),\\ \label{imsLA} &&G_{11}=g_{ab}Z'^a Z'^b +
2\Lambda^\mu_1 g_{\mu a}Z'^a + \Lambda^\mu_1\Lambda^\nu_1
g_{\mu\nu} ;\ea \ba\nl B_{01}=\Lambda^\mu_0\left(b_{\mu a}Z'^a +
\Lambda^\nu_1 b_{\mu\nu}\right);\ea

The constraints (\ref{0}) and (\ref{1}) respectively, and the
equations of motion for $X^M$ (\ref{sem}), can be written as
\ba\label{a0} &&\gamma^{11}\left(g_{ab}Z'^aZ'^b +
2\Lambda^\mu_1g_{\mu b}Z'^b\right) -
\left(\gamma^{00}\Lambda^\mu_0\Lambda^\nu_0 -
\gamma^{11}\Lambda^\mu_1\Lambda^\nu_1\right)g_{\mu\nu}=0,\\
\label{a1} &&\Lambda^\mu_0\left(\gamma^{11}g_{\mu a}Z'^a +
\gamma^{1n} \Lambda^\nu_n g_{\mu\nu}\right)=0;\ea \ba\label{aem}
&&\gamma^{11}\left(g_{Lb}Z''^b + \Gamma_{L,bc}Z'^bZ'^c \right) +
2\gamma^{1m}\Lambda^\mu_m\Gamma_{L,\mu b}Z'^b +
\gamma^{mn}\Lambda^{\mu}_{m}\Lambda^{\nu}_{n}\Gamma_{L,\mu\nu} \\
\nl &&=-\frac{1}{\sqrt{- \gamma}}\Lambda_0^\mu \left(H_{L\mu
a}Z'^a + \Lambda_1^\nu H_{L \mu \nu}\right).\ea

Let us write down the conserved quantities. By definition, the
generalized momenta are \ba \nl P_L\equiv\frac{\p
\mathcal{L}^P}{\p (\p_0 X^L)} = -T \left(\sqrt{-\gamma}\gamma
^{0n} g_{LN}\p_n X^N - b_{LN} \p_1 X^N \right). \ea For our ansatz
(\ref{sLA}) , they take the form: \ba\nl P_L = -T \sqrt{-
\gamma}\left[ \left(\gamma^{01} g_{Lb} - \frac{1}{\sqrt{-
\gamma}}b_{Lb}\right) Z'^b + \gamma ^{0n}\Lambda_n^\nu g_{L \nu}-
\frac{1}{\sqrt{- \gamma}}\Lambda ^\nu_1 b_{L \nu}\right].\ea The
Lagrangian (\ref{LRa}) does not depend on the coordinates $X^\mu$.
Therefore, the conjugated momenta $P_\mu$ do not depend on the
proper time $\tau$ \footnote{Actually, all momenta $P_M$ do not
depend on $\tau$, because there is no such dependence in
(\ref{LRa}).} \ba\label{cmla} P_\mu(\s) = -T \sqrt{- \gamma}
\left[ \left(\gamma^{01} g_{\mu b} - \frac{1}{\sqrt{-
\gamma}}b_{\mu b}\right) Z'^b+\gamma ^{0n}\Lambda_n^\nu g_{\mu
\nu}- \frac{1}{\sqrt{- \gamma}}\Lambda ^\nu_1 b_{\mu
\nu}\right],\hspace{.1cm} \p_0 P_\mu=0 .\ea

In order for our ansatz (\ref{sLA}), (\ref{ob}) to be  consistent
with the action (\ref{pa}), the following conditions must be
fulfilled \ba\label{emmu} \p_1\mathcal{P}_\mu\equiv
\frac{\p\mathcal{P}_\mu}{\p\s}=0,\ea where \ba\nl &&\mathcal{P}_M
\equiv\frac{\p \mathcal{L}^P}{\p (\p_1 X^M)} =-T\left(\sqrt{-
\gamma}\gamma ^{1n}g_{MN}\p_n X^N + b_{MN}\p_0 X^N\right)\\
\label{pM}&&=-T \sqrt{- \gamma}\left[ \gamma^{11} g_{Mb}Z'^b +
\gamma ^{1n}\Lambda_n^\nu g_{M\nu}+ \frac{1}{\sqrt{-
\gamma}}\Lambda ^\nu_0 b_{M\nu}\right].\ea This is because the
equations of motion (\ref{em}) can be rewritten as \ba\nl \frac{\p
P_M }{\p\tau} + \frac{\p \mathcal{P}_M}{\p\sigma} - \frac{\p
\mathcal{L}^P}{\p x^M}=0.\ea Hence, for the ansatz (\ref{sLA}),
(\ref{ob}), and for $M=\mu$, these equations take the form
(\ref{emmu}). Let us show this explicitly. In accordance with
(\ref{ob}), the computation of $\Gamma_{\lambda,MN}$ and
$H_{\lambda,MN}$ gives \ba\nl
&&\Gamma_{\lambda,ab}=\frac{1}{2}\left(\p_ag_{b\lambda}
+\p_bg_{a\lambda}\right), \h \Gamma_{\lambda,\mu
a}=\frac{1}{2}\p_ag_{\mu\lambda},\h \Gamma_{\lambda,\mu\nu}=0,\\
\nl &&H_{\lambda ab}=\p_a b_{b\lambda} + \p_b b_{\lambda a},\h
H_{\lambda\mu a}=\p_a b_{\lambda\mu},\h H_{\lambda\mu\nu}=0.\ea
Inserting these expressions in the part of the differential
equations (\ref{aem}) corresponding to $L=\lambda$, and using the
equalities $g'_{MN}=Z'^a\p_a g_{MN}$, $b'_{MN}=Z'^a\p_a b_{MN}$,
one receives that the quantities \ba\label{fi}
\gamma^{11}g_{\lambda a}(\s)Z'^a(\s) +
\gamma^{1m}\Lambda^{\mu}_{m}g_{\lambda\mu}(\s) + \frac{1}{\sqrt{-
\gamma}}\Lambda_0 ^\mu b_{\lambda\mu}(\s)\ea do not depend on
$\sigma$. Actually, they do not depend on $\tau$ too. Comparing
(\ref{fi}) with (\ref{pM}), we see that they are connected with
the constants of the motion $\mathcal{P}_\mu$ as \ba\label{cq}
\gamma^{11}g_{\mu a}Z'^a + \gamma^{1n}\Lambda^{\nu}_{n}g_{\mu\nu}
+ \frac{1}{\sqrt{- \gamma}}\Lambda_0 ^\nu b_{\mu\nu} =
-\frac{\mathcal{P}_\mu}{T\sqrt{-\gamma}} = constants.\ea

From (\ref{a1}) and (\ref{cq}), one obtains the following
compatibility condition \ba\label{cc}
\Lambda^{\nu}_{0}\mathcal{P}_\nu = 0.\ea This equality may be
interpreted as a solution of the constraint (\ref{a1}), which
restricts the number of the independent parameters in the theory
\footnote{When $\Lambda^{\mu}_{1}=0$, $g_{\mu a}=0$ and in
diagonal worldsheet gauge $(\gamma^{01}=0)$, the constraint
(\ref{a1}) and the condition (\ref{cc})  are identically
satisfied. All classical string configurations belonging to the
type (\ref{sLA}), and the corresponding string theory backgrounds
considered in \cite{4} - \cite{9}, \cite{12}, \cite{13},
\cite{27}, \cite{15}, \cite{19} - \cite{25}, \cite{29} -
\cite{33}, are particular cases of this particular case.}.

With the help of (\ref{cq}), the other constraint, (\ref{a0}), can
be rewritten in the form \ba\label{ec} g_{ab}Z'^aZ'^b =
\mathcal{U},\ea where $\mathcal{U}$ is given by \ba\label{sp}
\mathcal{U}=\frac{1}{\gamma^{11}} \left[
\gamma^{mn}\Lambda^{\mu}_{m}\Lambda^{\nu}_{n}g_{\mu\nu} +
\frac{2\Lambda^{\mu}_{1}}{T\sqrt{-\gamma}} \left(\mathcal{P}_\mu+
T \Lambda_0^\nu b_{\mu \nu}\right) \right].\ea

Now, let us turn to the equations of motion (\ref{aem}),
corresponding to $L=a$. In view of the conditions (\ref{ob}),
\ba\nl &&\Gamma_{a,\mu
b}=-\frac{1}{2}\left(\p_ag_{b\mu}-\p_bg_{a\mu}\right)
=-\p_{[a}g_{b]\mu},\h
\Gamma_{a,\mu\nu}=-\frac{1}{2}\p_ag_{\mu\nu},\\ \nl &&
H_{a\mu\nu}= \p_a b_{\mu\nu}; \h H_{ab\nu}=\p_ab_{b\nu} - \p_b
b_{a \nu}= 2\p_{[a}b_{b]\nu}.\ea By using this, one obtains
\ba\label{fem} g_{ab}Z''^b + \Gamma_{a,bc}Z'^bZ'^c =
\frac{1}{2}\p_a \mathcal{U} + 2\p_{[a}\mathcal{A}_{b]}Z'^b.\ea In
(\ref{fem}), an effective scalar potential $\mathcal{U}$ and an
effective 1-form gauge field $\mathcal{A}_a$ appeared.
$\mathcal{U}$ is given in (\ref{sp}) (and is the same as in the
effective constraint (\ref{ec})), and \ba\label{gf} \mathcal{A}_a=
\frac{1}{\gamma ^{11}}\left(\gamma^{1 m}\Lambda_m^\mu g_{a\mu}+
\frac{1}{\sqrt{-\gamma}}\Lambda_0^\mu b_{a \mu} \right).\ea

Now our task is to find {\it exact} solutions of the {\it
nonlinear} differential equations (\ref{ec}) and (\ref{fem}).
How exactly this can be done is explained in Appendix A.
Here, we give the final results only.

If the background seen by the string depends on only one
coordinate $x^a$, the general solution for the string embedding
coordinate $X^a(\tau,\s)=Z^a(\s)$ is given by
\ba\nl\sigma\left(X^a\right)=\sigma_0 + \int_{X_0^a}^{X^a}
\left(\frac{\mathcal{U}}{g_{aa}}\right)^{-1/2}dx .\ea When the
background felt by the string depends on more than one coordinate
$x^a$, the first integrals of the equations of motion for
$Z^a(\s)=(Z^r, Z^\alpha)$, which also solve the constraint
(\ref{ec}), are \ba\nl &&\left(g_{rr}Z'^r\right)^2 = g_{rr}
\left[(1-n_\alpha)\mathcal{U} -2n_\alpha\left(\mathcal{A}_r-\p_r
f\right)Z'^r -\sum_{\alpha}\frac{D_{\alpha}
\left(Z^{a\ne\alpha}\right)} {g_{\alpha\alpha}}\right]=
F_r\left(Z^r\right)\ge 0,\\ \nl
&&\left(g_{\alpha\alpha}Z'^\alpha\right)^2 =D_{\alpha}
\left(Z^{a\ne\alpha}\right) + g_{\alpha\alpha}\left[\mathcal{U}
+2\left(\mathcal{A}_r-\p_r f\right)Z'^r\right]= F_{\alpha}
\left(Z^{\beta}\right)\ge 0,\ea where $Z^r$ is one of the
coordinates $Z^a$, $Z^\alpha$ are the remaining ones, $n_\alpha$
is the number of $Z^\alpha$, and $D_{\alpha}$, $F_{a}$ are
arbitrary functions of their arguments. The above expressions are
valid, if the $g_{ab}$ part of the metric is diagonal one, and the
following integrability conditions hold\footnote{In all cases,
considered in \cite{4} - \cite{34}, $\mathcal{A}_a\equiv 0$.}
\ba\nl &&\mathcal{A}_a \equiv(\mathcal{A}_r,\mathcal{A}_\alpha)=
(\mathcal{A}_r,\p_\alpha f),\h
\p_\alpha\left(\frac{g_{\alpha\alpha}}{g_{aa}}\right)=0,
\\ \nl &&\p_\alpha\left(g_{rr}Z'^r\right)^2=0,\h
\p_r\left(g_{\alpha\alpha}Z'^\alpha\right)^2=0. \ea

\subsection{The tensionless limit}
The results obtained till now are not applicable to tensionless
(null) strings, because the action (\ref{pa}) is proportional to
the string tension $T$. The parameterization of the auxiliary
worldsheet metric $\gamma^{mn}$, which is appropriate for
considering the zero tension limit $T\to 0$, is the following
\cite{ILST93, HLU94}: \ba\label{tl} \gamma^{00}=-1,\h
\gamma^{01}=\lambda^1,\h \gamma^{11}=(2\lambda^0T)^2 -
(\lambda^1)^2, \h \det(\gamma^{mn})= -(2\lambda^0T)^2.\ea Now, the
action (\ref{pa}) becomes  \ba\nl S_\lambda =\int
d^2\xi\Bigl\{\frac{1}{4\lambda^0}\Bigl[
G_{00}-2\lambda^{1}G_{01}+\left(\lambda^{1}\right)^2 G_{11}
-\left(2\lambda^0T\right)^2 G_{11}\Bigr] + TB_{01}\Bigr\}.\ea
Here, $\lambda^n$ are the Lagrange multipliers, whose equations of
motion generate the {\it independent} constraints.

In these notations, the constraints (\ref{a0}) and (\ref{a1}), the
equations of motion (\ref{aem}), and the conserved quantities
(\ref{cmla}), (\ref{cq}) take the form \ba\nl &&g_{ab}Z'^aZ'^b +
2\Lambda_{1}^{\mu}g_{\mu b}Z'^b +
\left[\frac{\Lambda^\mu_0\Lambda^\nu_0}{(2\lambda^0T)^2 -
(\lambda^1)^2}+\Lambda^\mu_1\Lambda^\nu_1\right]g_{\mu\nu}=0,\\
\nl &&\Lambda^\mu_0\left\{g_{\mu a}Z'^a +
\left[\frac{\lambda^1\Lambda^\nu_0}{(2\lambda^0T)^2 -
(\lambda^1)^2}+\Lambda^\nu_1\right]g_{\mu\nu}\right\}=0;\ea \ba\nl
&&g_{Lb}Z''^b + \Gamma_{L,bc}Z'^bZ'^c +
2\left[\frac{\lambda^1\Lambda^\mu_0}{(2\lambda^0T)^2 -
(\lambda^1)^2}+\Lambda^\mu_1\right]\Gamma_{L,\mu b}Z'^b\\ \nl &&+
\left[\frac{\Lambda^\mu_0\left(2\lambda^1\Lambda^\nu_1 -
\Lambda^\nu_0 \right)}{(2\lambda^0T)^2 - (\lambda^1)^2}
+\Lambda^\mu_1\Lambda^\nu_1\right]\Gamma_{L,\mu\nu}
=-\frac{2\lambda^0 T}{(2\lambda^0T)^2 -
(\lambda^1)^2}\Lambda_0^\mu \left(H_{L\mu a}Z'^a + \Lambda_1^\nu
H_{L \mu \nu}\right).\ea \ba\nl &&P_\mu(\s)
=\frac{1}{2\lambda^0}\left[\left(-\lambda^1g_{\mu a} +
2\lambda^0Tb_{\mu a}\right)Z'^a +
\left(\Lambda^{\nu}_0-\lambda^1\Lambda^{\nu}_1\right)g_{\mu\nu} +
2\lambda^0T\Lambda^{\nu}_1b_{\mu\nu}\right],\\ \nl
&&\mathcal{P}_{\mu} =-\frac{1}{2\lambda^0}
\left\{\left[(2\lambda^0T)^2-(\lambda^1)^2\right]\left(g_{\mu a}
Z'^{a}+ \Lambda^{\nu}_1 g_{\mu\nu}\right)+ \Lambda^{\nu}_0
\left(\lambda^1 g_{\mu\nu} + 2\lambda^0 T
b_{\mu\nu}\right)\right\}.\ea

The reduced equations of motion and constraint (\ref{fem}) and
(\ref{ec}) have the same form, but now, the effective potential
(\ref{sp}) and the effective gauge field (\ref{gf}) are given by
\ba\nl &&\mathcal{U} =
\left[\frac{\Lambda^\mu_0\left(2\lambda^1\Lambda^\nu_1 -
\Lambda^\nu_0 \right)}{(2\lambda^0T)^2 - (\lambda^1)^2}
+\Lambda^\mu_1\Lambda^\nu_1\right]g_{\mu\nu} +
\frac{4\lambda^0}{(2\lambda^0T)^2 -
(\lambda^1)^2}\Lambda^\mu_1\left( \mathcal{P}_\mu + T\Lambda^\nu_0
b_{\mu\nu}\right),\\ \nl &&\mathcal{A}_a =
\left[\frac{\lambda^1\Lambda^\nu_0}{(2\lambda^0T)^2 -
(\lambda^1)^2}+\Lambda^\nu_1\right]g_{a\nu} + \frac{2\lambda^0
T}{(2\lambda^0T)^2 - (\lambda^1)^2}\Lambda_0^\mu b_{a\mu}.\ea

If one sets $\lambda^1=0$ and $2\lambda^0T=1$, this will
correspond to {\it conformal gauge}, as it should be. If one puts
$T=0$ in the above formulas, they will describe {\it tensionless}
strings.

\setcounter{equation}{0}
\section{Exact solutions for more general string embedding}
In this section, we will use the ansatz (\ref{sGA}) for the string
coordinates, which corresponds to more general string embedding.
Here, compared with (\ref{sLA}), $X^\mu$ are allowed to vary
non-linearly with the worldsheet spatial parameter $\sigma$. Of
course, the conditions (\ref{ob}) on the background fields are
also fulfilled.

Taking into account the ansatz (\ref{sGA}), one obtains that the
induced fields $G_{mn}$ and $B_{mn}$, the Lagrangian density, the
constraints (\ref{0}) and (\ref{1}) respectively, and the
Euler-Lagrange equations for $X^M$ (\ref{sem}) are given by \ba
\label{imga} &&G_{00}=\Lambda^\mu_0\Lambda^\nu_0 g_{\mu\nu},\h
G_{01}=\Lambda^\mu_0\left(g_{\mu N}Z'^N + \Lambda^\nu_1
g_{\mu\nu}\right),\\ \nl &&G_{11}=g_{MN}Z'^M Z'^N + 2\Lambda^\mu_1
g_{\mu N}Z'^N + \Lambda^\mu_1\Lambda^\nu_1 g_{\mu\nu} ;\ea \ba\nl
B_{01}=\Lambda^\mu_0\left(b_{\mu N}Z'^N + \Lambda^\nu_1
b_{\mu\nu}\right);\ea \ba\nl\mathcal{L}^{GA}(\sigma) =
-\frac{T}{2}\sqrt{-\gamma}\left[\gamma^{11}g_{MN}Z'^MZ'^N + 2
\left(\gamma^{1m}\Lambda_m^\mu g_{\mu N}- \frac{\Lambda_0^\mu b_{
\mu N}}{\sqrt{- \gamma}} \right)Z'^N + \right.\\ \nl +\left.
\gamma^{mn}\Lambda_m^\mu \Lambda_n^\nu g_{\mu \nu}- \frac{2
\Lambda_0^\mu\Lambda_1^\nu b_{\mu\nu}}{\sqrt{-\gamma}} \right];\ea
\ba\label{mga0} &&\gamma^{11}g_{MN}Z'^M Z'^N +
2\gamma^{11}\Lambda^\mu_1g_{\mu N}Z'^N -
\left(\gamma^{00}\Lambda^\mu_0\Lambda^\nu_0 -
\gamma^{11}\Lambda^\mu_1\Lambda^\nu_1\right)g_{\mu\nu}=0,\\
\label{mga1} &&\Lambda^\mu_0\left(\gamma^{11}g_{\mu N}Z'^N +
\gamma^{1n} \Lambda^\nu_n g_{\mu\nu}\right)=0;\ea \ba\label{mgaem}
\gamma^{11}\left(g_{LN}Z''^N + \Gamma_{L,MN}Z'^M Z'^N \right) +
2\gamma^{1m}\Lambda^\mu_m\Gamma_{L,\mu N}Z'^N +
\gamma^{mn}\Lambda^{\mu}_{m}\Lambda^{\nu}_{n}\Gamma_{L,\mu\nu}= \\
\nl = -\frac{1}{\sqrt{- \gamma}}\Lambda_0 ^\mu\left(H_{L\mu N}
Z'^N + \Lambda_1^\nu H_{L \mu\nu}\right) .\ea The quantities
$P_L$, $\mathcal{P}_L$ can be found as before, and now they are
\ba\label{mgcm} &&\left(\gamma^{01}g_{L N} - \frac{b_{LN}}{\sqrt{-
\gamma}}\right)Z'^N + \gamma^{0n}\Lambda^{\nu}_{n}g_{L\nu}-
\frac{\Lambda_1^\nu b_{L\nu}}{\sqrt{- \gamma}} =
-\frac{P_L}{T\sqrt{-\gamma}},\h \p_0 P_L=0,\\ \label{mgcq} &&
\gamma^{11}g_{LN}Z'^N + \gamma^{1n}\Lambda^{\nu}_{n}g_{L\nu} +
\frac{\Lambda_0 ^\nu b_{L\nu}}{\sqrt{- \gamma}} =
-\frac{\mathcal{P}_L}{T\sqrt{-\gamma}},\h \p_0 \mathcal{P}_L=0, \h
\p_1 \mathcal{P}_\mu=0.\ea The compatibility condition following
from the constraint (\ref{mga1}) and from (\ref{mgcq}) coincides
with the previous one (\ref{cc}).

As in the previous section, the equations (\ref{mgaem}) for
$L=\lambda$ lead to $\p_1\mathcal{P}_{\lambda}=0$. Consequently,
our next task is to consider the equations (\ref{mgaem}) for $L=a$
and the constraint (\ref{mga0}). First of all, we will eliminate
the variables $Z'^\mu$ from them. To this end, we express $Z'^\mu$
through $Z'^a$ by using (\ref{mgcq}): \ba\label{muc} Z'^\mu =
-\frac{\gamma^{1m}}{\gamma^{11}}\Lambda^\mu_m -
\left(g^{-1}\right)^{\mu\nu}\left[g_{\nu a}Z'^a +
\frac{1}{T\sqrt{-\gamma}\gamma^{11}}\left( \mathcal{P}_\nu +
T\Lambda_0^\rho b_{\nu \rho} \right)\right].\ea With the help of
(\ref{muc}) and (\ref{cc}), the equations (\ref{mgaem}) for $L=a$
and the constraint (\ref{mga0}) acquire the form \ba\label{mgemf}
&&h_{ab}Z''^b + \Gamma^{\bf{h}}_{a,bc}Z'^b Z'^c = \frac{1}{2}\p_a
\mathcal{U}^{\bf{h}} + 2\p_{[a}\mathcal{A}^{\bf{h}}_{b]}Z'^b,\\
\label{mgecf} &&h_{ab}Z'^aZ'^b = \mathcal{U}^{\bf{h}},\ea where a
new, effective metric appeared \ba\nl h_{ab} = g_{ab} -
g_{a\mu}(g^{-1})^{\mu\nu}g_{\nu b}.\ea $\Gamma^{\bf{h}}_{a,bc}$ is
the symmetric connection corresponding to this metric \ba\nl
\Gamma^{\bf{h}}_{a,bc}=\frac{1}{2}\left(\p_bh_{ca}
+\p_ch_{ba}-\p_ah_{bc}\right).\ea The effective scalar and gauge
potentials, expressed through the background fields, are as
follows \ba\nl
&&\mathcal{U}^{\bf{h}}=\frac{1}{\gamma\left(\gamma^{11}\right)^2}
\left[ \Lambda^{\mu}_{0}\Lambda^{\nu}_{0}g_{\mu\nu} +
\frac{1}{T^2} \left( \mathcal{P}_\mu + T\Lambda_0^\rho b_{\mu
\rho}\right)(g^{-1})^{\mu\nu} \left( \mathcal{P}_\nu +
T\Lambda_0^\lambda b_{\nu\lambda}\right)\right],
\\ \nl &&\mathcal{A}^{\bf{h}}_{a}= -
\frac{1}{T\sqrt{-\gamma}\gamma^{11}} \left[g_{a\mu}
(g^{-1})^{\mu\nu} \left(\mathcal{P}_\nu + T\Lambda _0 ^ \rho
b_{\nu\rho} \right)- T \Lambda_0 ^\rho b_{a \rho} \right] .\ea We
point out the qualitatively different behavior of the potentials
$\mathcal{U}^{\bf{h}}$ and $\mathcal{A}_a^{\bf{h}}$, compared to
$\mathcal{U}$ and $\mathcal{A}_a$ from the previous section, due
to the appearance of the inverse metric $(g^{-1})^{\mu\nu}$ in the
above expressions.

Since the equations (\ref{fem}), (\ref{ec}) and (\ref{mgemf}),
(\ref{mgecf}) have the same form, for obtaining exact string
solutions, we  can proceed as before and use the derived formulas
after the replacements $(g,\Gamma,\mathcal{U},\mathcal{A})$ $\to$
$(h,\Gamma^{\bf{h}}, \mathcal{U}^{\bf{h}},\mathcal{A}^{\bf{h}})$
(see Appendix A). In particular, the solution depending on one of
the coordinates $X^a$ will be
\ba\label{mgocs}\sigma\left(X^a\right)=\sigma_0 +
\int_{X_0^a}^{X^a}d x
\left(\frac{\mathcal{U}^{\bf{h}}}{h_{aa}}\right)^{-1/2}.\ea In
this case by integrating (\ref{muc}), and replacing the solution
for $Z^\mu$ in the ansatz (\ref{sGA}), one obtains solution for
the string coordinates $X^\mu$ of the type $X^\mu(\tau, X^a)$: \ba
\label{X} &&X^\mu(\tau, X^a) = X_0^\mu + \Lambda_0 ^\mu \left[
\tau  - \frac{\gamma^{01}}{\gamma^{11}}
\sigma\left(X^a\right)\right] -\\ \nl && - \int_ {X_0^a}^ {X^a}
(g^{-1})^{\mu\nu} \left[g_{\nu a}+ \frac{\left(\mathcal{P}_\nu +
T\Lambda_0^\rho b_{\nu\rho}\right)} {T \sqrt{- \gamma}\gamma^{11}}
\left(\frac{\mathcal {U}^{\rm{h}}}{h_{aa}} \right)^{-1/2} \right]
d x .\ea To write down a solution of the type $X^\mu(\tau,\s)$,
one have to invert the solution (\ref{mgocs}):
$\sigma\left(X^a\right)\to X^a(\s)$. Then, $X^\mu(\tau,\s)$ are
given by \ba\label{aX} &&X^\mu(\tau, \s) = X_0^\mu + \Lambda_0
^\mu \left(\tau -\frac{\gamma^{01}}{\gamma^{11}} \sigma\right) -\\
\nl && - \int_{\s_0}^{\s} (g^{-1})^{\mu\nu}
\left[\frac{\left(\mathcal{P}_\nu + T\Lambda_0^\rho
b_{\nu\rho}\right)} {T \sqrt{- \gamma}\gamma^{11}} +g_{\nu a}
\left(\frac{\mathcal {U}^{\rm{h}}}{h_{aa}} \right)^{1/2}\right]
d\s .\ea

Let us also give the expression for $P_\mu$ after the elimination
of $Z'^\mu$ from (\ref{mgcm}) \ba\label{cmcq}
&&P_\mu(\s)=T\left[b_{\mu
a}-b_{\mu\nu}\left(g^{-1}\right)^{\nu\lambda}g_{\lambda a
}\right]Z'^a\\ \nl
&&+\frac{1}{\gamma^{11}}\left\{\gamma^{01}\mathcal{P}_{\mu} +
\frac{1}{\sqrt{-\gamma}}\left[T\Lambda^\nu_0g_{\mu\nu} -
b_{\mu\nu}\left(g^{-1}\right)^{\nu\lambda}\left(
\mathcal{P}_\lambda+T\Lambda^\rho_0b_{\lambda\rho}\right)\right]
\right\}.\ea These equalities connect the conserved momenta
$P_\mu$ with the constants of the motion $\mathcal{P}_{\mu}$.

To be able to take the tensionless limit $T\to 0$ in the above
formulas, we must use the $\lambda$-parameterization (\ref{tl}) of
$\gamma^{mn}$. The quantities, which depend on this
parameterization, and appear in the reduced equations of motion
and constraint (\ref{mgemf}), (\ref{mgecf}), and therefore - in
the solutions, are $\mathcal{U}^{\bf{h}}$ and
$\mathcal{A}^{\bf{h}}_{a}$. Now, they read \ba\nl
&&\mathcal{U}^{\bf{h}} =
-\frac{(2\lambda^0)^2}{\left[(2\lambda^0T)^2-(\lambda^1)^2\right]^2}
\left[ T^2\Lambda^{\mu}_{0}\Lambda^{\nu}_{0}g_{\mu\nu} + \left(
\mathcal{P}_\mu + T\Lambda_0^\rho b_{\mu
\rho}\right)(g^{-1})^{\mu\nu} \left( \mathcal{P}_\nu +
T\Lambda_0^\lambda b_{\nu\lambda}\right)\right],
\\ \nl &&\mathcal{A}^{\bf{h}}_{a}= -
\frac{2\lambda^0}{(2\lambda^0T)^2-(\lambda^1)^2} \left[g_{a\mu}
(g^{-1})^{\mu\nu} \left(\mathcal{P}_\nu + T\Lambda _0 ^ \rho
b_{\nu\rho} \right)- T \Lambda_0 ^\rho b_{a \rho} \right] .\ea If
one sets $\lambda^1=0$ and $2\lambda^0T=1$, the {\it conformal
gauge} results are obtained. If one puts $T=0$ in the above
equalities, they will correspond to {\it tensionless} strings. For
instance, the solution $X^\mu(\tau, X^a)$ reduces to \ba \nl
&&X^\mu(\tau, X^a)_{T=0} = X_0^\mu + \Lambda_0 ^\mu \left[ \tau +
\frac{\sigma\left(X^a\right)}{\lambda^1}\right] -\\ \nl && - \int_
{X_0^a}^ {X^a} (g^{-1})^{\mu\nu} \left[g_{\nu a}-
\frac{2\lambda^0}{(\lambda^1)^2}\mathcal{P}_\nu
\left(\frac{\mathcal {U}^{\rm{h}}}{h_{aa}} \right)^{-1/2}_{T=0}
\right] d x .\ea

\setcounter{equation}{0}
\section{Examples}

In the previous two sections, we described a general approach for
solving the string equations of motion and constraints in the
background fields $g_{MN}(x)$ and $b_{MN}(x)$, based on the
ansatzes (\ref{sLA}) and (\ref{sGA}). In this section, as an
illustration of the previously obtained generic results, we will
establish the correspondence with the particular cases considered
in \cite{4} - in the framework of the linear ansatz (\ref{sLA}),
and in \cite{17} - in the framework of the nonlinear ansatz
(\ref{sGA}). After that, we will find new exact string solutions
in $AdS_5$ black hole and in $AdS_5\times S^5$ backgrounds.

\subsection{Comparing with known solutions}

In \cite{4}, the string theory background is $AdS_5\times S^5$,
with field theory dual $\mathcal{N}=4$ $SU(N)$ $SYM$ in four
dimensional flat space-time. The metric of the  $AdS_5$ is taken
in global coordinates, so that the string energy is identified
with the conformal dimension in the dual CFT \ba\nl
&&ds^2_{AdS_5}= R^2\left(-\cosh^2\rho dt^2 + d\rho^2 + \sinh^2\rho
d\Omega^2_3\right),\\ \nl &&d\Omega^2_3= d\theta^2 +
\cos^2\theta\left(d\psi^2 + \cos^2\psi d\phi^2\right),\h
R^4=\lambda\alpha'^2.\ea

As our first example, we will consider one of the classical string
configurations analyzed in \cite{4}. Namely, we will consider a
closed spinning string in $AdS_5$ given by the following ansatz
\ba\label{pa1} t=e\tau,\h \rho=\rho(\sigma),\h \phi=e\omega\tau,\h
e, \omega = constants .\ea Therefore, the background seen by the
string is \ba\label{rm1} ds^2=R^2\left(-\cosh^2\rho dt^2 + d\rho^2
+ \sinh^2\rho d\phi^2\right), \h b_{MN}=0.\ea This metric does not
depend on $x^0=t$ and $x^2=\phi$, i.e. $X^\mu=X^{0,2}$ and
$X^a=X^1=\rho(\s)$ in our notation. Comparing the ansatzes
(\ref{sLA}) and (\ref{pa1}), one sees that the latter is
particular case of the former, corresponding to \ba\nl
\Lambda_0^0=e,\h \Lambda^0_1=\Lambda_1^2=0,\h
\Lambda_0^2=e\omega.\ea

The conserved momenta $P_\mu$, obtained from (\ref{cmla}), are
\ba\nl P_0 = T\sqrt{-\gamma}\gamma^{00}eR^2\cosh^2\rho,\h P_2 = -
T\sqrt{-\gamma}\gamma^{00}e\omega R^2\sinh^2\rho.\ea So, the
energy and the spin are given by \ba\label{ce1}
&&E=-\int_{0}^{2\pi}d\s P_0 = -T\sqrt{-\gamma}\gamma^{00}eR^2
\int_{0}^{2\pi}d\s\cosh^2\rho, \\ \label{cs1}
&&S=\int_{0}^{2\pi}d\s P_2 = - T\sqrt{-\gamma}\gamma^{00}e\omega
R^2 \int_{0}^{2\pi}d\s\sinh^2\rho.\ea

In diagonal worldsheet gauge, $\gamma^{01}=0$, the first integrals
(\ref{cq}) are identically zero: $\mathcal{P}_\mu\equiv 0$, $\mu=0,2$. In
this gauge, the constraint (\ref{a1}) and the compatibility condition
(\ref{cc}) are also identically satisfied. Thus, it remains to solve the
equations of motion (\ref{fem}) and the constraint (\ref{ec}). The
background (\ref{rm1}) depends on only one coordinate $x^1=\rho$. We know
from our considerations in Appendix A that in this case, the effective
constraint (\ref{ec}) is first integral of the equation (\ref{fem}). It
reads \ba\label{ece1} \rho'^2= -\frac{\gamma^{00}}{\gamma^{11}}e^2
\left(\cosh^2\rho-\omega^2\sinh^2\rho\right).\ea It follows from here that
\ba\label{esdf} d\s = \frac{d\rho}{e\sqrt{
-\frac{\gamma^{00}}{\gamma^{11}}
\left(\cosh^2\rho-\omega^2\sinh^2\rho\right)}}.\ea By integrating this
equality, one obtains the string solution $\s(\rho)$, given in the general
case by (\ref{tsol1}).

In {\it conformal gauge}, (\ref{ce1}), (\ref{cs1}), (\ref{ece1})
and (\ref{esdf}) reduce to the corresponding formulas derived in
\cite{4}. In the {\it tensionless limit}, after using the
$\lambda$-parameterization (\ref{tl}) of $\gamma^{mn}$,
(\ref{ce1}), (\ref{cs1}) and (\ref{esdf}) take the form \ba\nl
&&E_{T=0}= \frac{eR^2}{2\lambda^0}\int_{0}^{2\pi}d\s\cosh^2\rho,
\\ \nl &&S_{T=0}= \frac{e\omega R^2}{2\lambda^0}
\int_{0}^{2\pi}d\s\sinh^2\rho, \\ \nl &&d\s_{T=0} =
\frac{\lambda^1 d\rho}{e\sqrt{
\left(\omega^2\sinh^2\rho-\cosh^2\rho\right)}}.\ea

In \cite{17}, the string theory background is $AdS_3\times
S^3\times\mathcal{M}$, with NS-NS 2-form gauge field. In
accordance with the $AdS/CFT$ duality, the string theory on
$AdS_3\times S^3\times\mathcal{M}$ is dual to a superconformal
field theory on a cylinder, which is the boundary of $AdS_3$ in
global coordinates. In such coordinates, the NS-NS $AdS_3$
background can be written as \ba\label{rm2} ds^2_{AdS_3} =
R^2\left[-(1+r^2)dt^2 + \frac{dr^2}{1+r^2} + r^2 d\phi^2\right],\h
b_{t\phi}=R^2 r^2.\ea

As our second example, we will consider the first of the two
classical string configurations discussed in \cite{17}
\ba\label{pa2} t=c_1\tau+\tilde{t}(\s),\h r=r(\s),\h
\phi=c_2\tau+\tilde{\phi}(\s).\ea The above background does not
depend on $x^0=t$ and $x^2=\phi$. Therefore, $\mu=(0,2)$, $a=1$,
and comparing (\ref{pa2}) with (\ref{sGA}), one obtains \ba\nl
&&\Lambda^0_0=c_1,\h \Lambda^2_0=c_2,\h \Lambda^\mu_1=0,\\ \nl
&&Z^0(\s)=\tilde{t}(\s),\h Z^1(\s)=r(\s),\h
Z^2(\s)=\tilde{\phi}(\s).\ea The action (\ref{pa}) for this string
embedding, in worldsheet gauge $\gamma^{mn}=constants$, is
\ba\label{ae2} &&S^P=-\frac{TR^2}{2}\sqrt{-\gamma}\int d\tau
d\sigma\left[-(1+r^2)\left(\gamma^{00}c_1^2 + 2\gamma^{01}c_1
\tilde{t}' + \gamma^{11}\tilde{t}'^2\right)\right. \\ \nl  &&+
\gamma^{11}\frac{r'^2}{1+r^2} + \left. r^2\left(\gamma^{00}c_2^2 +
2\gamma^{01}c_2 \tilde{\phi}' + \gamma^{11}\tilde{\phi}'^2\right)
- \frac{2}{\sqrt{-\gamma}}r^2\left(c_1\tilde{\phi}' -
c_2\tilde{t}'\right) \right].\ea

The explicit expressions for the conserved momenta $P_\mu$ can be
obtained from (\ref{mgcm}). By using them, one is able to compute
the string energy and spin: \ba\label{ce2} &&E=-\int_{0}^{2\pi}d\s
P_0 = -T\sqrt{-\gamma}R^2
\int_{0}^{2\pi}d\s\left[\left(\gamma^{00}c_1 +
\gamma^{01}t'\right)(1+r^2) +
\frac{r^2\phi'}{\sqrt{-\gamma}}\right],
\\ \label{cs2} &&S=\int_{0}^{2\pi}d\s P_2 = -
T\sqrt{-\gamma}R^2 \int_{0}^{2\pi}d\s\left(\gamma^{00}c_2 +
\gamma^{01}\phi' + \frac{t'}{\sqrt{-\gamma}}\right)r^2.\ea

As we know from Sec.3, the quantities $\mathcal{P}_\mu$, given in
(\ref{mgcq}), are first integrals for this part of the equations
of motion (\ref{mgaem}), which  correspond to $L=\mu$. For the
case under consideration, (\ref{mgcq}) leads to the following
expressions for $\mathcal{P}_\mu$: \ba\nl &&\mathcal{P}_0 =
T\sqrt{-\gamma}R^2\left[\left(\gamma^{11}\tilde{t}' +
\gamma^{01}c_1 \right)(1+r^2) -
\frac{c_2r^2}{\sqrt{-\gamma}}\right]=constant,\\ \nl
&&\mathcal{P}_2 =
-T\sqrt{-\gamma}R^2\left(\gamma^{11}\tilde{\phi}' + \gamma^{01}c_2
- \frac{c_1}{\sqrt{-\gamma}}\right)r^2=constant.\ea The above
equalities are first order ordinary differential equations for
$\tilde{t}(\s)$ and $\tilde{\phi}(\s)$, which rewritten in normal
form read \ba\label{fot}
\tilde{t}'=\frac{c_2r^2-k_1}{\sqrt{-\gamma}\gamma^{11}(1+r^2)}
-\frac{\gamma^{01}}{\gamma^{11}}c_1,\h
\tilde{\phi}'=\frac{c_1r^2-k_2}{\sqrt{-\gamma}\gamma^{11}r^2}
-\frac{\gamma^{01}}{\gamma^{11}}c_2,\ea where \ba\nl k_1\equiv
-\frac{\mathcal{P}_0}{TR^2},\h k_2\equiv
\frac{\mathcal{P}_2}{TR^2}.\ea The equations (\ref{fot}) can be
solved, if the function $r=r(\s)$ is already known. Before
explaining how to find it, let us consider the compatibility
condition (\ref{cc}), which is the same for both ansatzes -
(\ref{sLA}) and (\ref{sGA}). In our case, it gives \ba\label{cc2}
c_1k_1 = c_2k_2,\ea from where one can eliminate one of the free
parameters in the theory. From now on, we will use that
$c_2=c_1k_1/k_2$.

Now, we are going to solve the equations of motion (\ref{mgemf})
and the constraint (\ref{mgecf}). The background fields in
(\ref{rm2}) depend on only one coordinate $x^1=r$. Therefore, as
we already know, the effective constraint (\ref{mgecf}) is first
integral of the equation of motion (\ref{mgemf}) for this
coordinate.Thus, we have to only solve the constraint
(\ref{mgecf}), which in the case under consideration reads
\ba\label{ece2} r'^2=
-\frac{\alpha\beta}{\gamma\left(\gamma^{11}\right)^2 k_2^2 r^2}
\left(\frac{k_2^2}{\alpha}-r^2\right)
\left(r^2-\frac{k_2^2}{\beta}\right),\ea where \ba\nl
\alpha=k_1^2-k_2^2,\h \beta=c_1(c_1 + 2k_2).\ea To find solution
of this equation, we apply our general formula (\ref{mgocs}) and
obtain ($\s_0=0$): \ba\nl &&\s(r)=
\frac{\sqrt{-\gamma}\gamma^{11}k_2}{2\sqrt{\alpha\beta}}
\int\frac{dr^2}{\sqrt{\left(\frac{k_2^2}{\alpha}-r^2\right)
\left(r^2-\frac{k_2^2}{\beta}\right)}}\\ \nl &&=
-\sqrt{-\gamma}\gamma^{11}\frac{k_2}{2\sqrt{\alpha\beta}}\arcsin\left[
\frac{\frac{2\alpha\beta}{k_2^2}r^2-(\alpha+\beta)}{\alpha-\beta}
\right].\ea By inverting this solution, one receives
\ba\label{isr}
r^2(\s)=\frac{k_2^2}{2\alpha\beta}\left[(\alpha+\beta) -
(\alpha-\beta)\sin\left(
\frac{2\sqrt{\alpha\beta}}{\sqrt{-\gamma}\gamma^{11}k_2}\s\right)
\right].\ea

Now, we can write down the solutions $X^\mu=X^\mu(\tau,r)$ for the
remaining string coordinates, by using (\ref{X}): \ba\nl
&&X^0(\tau,r)\equiv t(\tau,r) = c_1\tau +
\frac{c_1}{\gamma^{11}}\left( \frac{k_1}{\sqrt{-\gamma}k_2} -
\gamma^{01}\right)\s(r)\\ \nl &&-
\frac{k_1(c_1+k_2)}{2\sqrt{\alpha\beta}}
\int\frac{dr^2}{(r^2+1)\sqrt{\left(\frac{k_2^2}{\alpha}-r^2\right)
\left(r^2-\frac{k_2^2}{\beta}\right)}},\\ \nl &&X^2(\tau,r)\equiv
\phi(\tau,r) = c_1\frac{k_1}{k_2}\tau +
\frac{c_1}{\gamma^{11}}\left( \frac{1}{\sqrt{-\gamma}} -
\gamma^{01}\frac{k_1}{k_2}\right)\s(r)\\ \nl &&-
\frac{k^2_2}{2\sqrt{\alpha\beta}}
\int\frac{dr^2}{r^2\sqrt{\left(\frac{k_2^2}{\alpha}-r^2\right)
\left(r^2-\frac{k_2^2}{\beta}\right)}}.\ea

Let us also give the connection between the conserved momenta
$P_\mu$ and the constants of the motion $\mathcal{P}_\mu$
following from (\ref{cmcq}): \ba\nl &&P_0=
\frac{1}{\gamma^{11}}\left[\gamma^{01}\mathcal{P}_0 -
\frac{1}{\sqrt{-\gamma}}\left( \mathcal{P}_2 +
c_1TR^2\right)\right],\\ \nl &&P_2=
\frac{1}{\gamma^{11}}\left[\gamma^{01}\mathcal{P}_2 -
\frac{1}{\sqrt{-\gamma}}\left( \mathcal{P}_0 -\frac{\mathcal{P}_0
+ c_2TR^2 r^2}{r^2+1}\right)\right].\ea By integrating these
equalities, one obtains the following expressions for the string
energy (\ref{ce2}) and spin (\ref{cs2}) \ba\label{se2} &&E=-
\frac{2\pi}{\gamma^{11}}\left[\gamma^{01}\mathcal{P}_0 - \frac{
\mathcal{P}_2}{\sqrt{-\gamma}}\left(1 + c_1\frac{TR^2}{
\mathcal{P}_2}\right)\right],\\ \label{ss2} &&S=
\frac{2\pi}{\gamma^{11}}\left[\gamma^{01}\mathcal{P}_2 - \frac{
\mathcal{P}_0}{\sqrt{-\gamma}}\left(1 + c_1\frac{TR^2}{
\mathcal{P}_2}\right)\left(1-\frac{1}{2\pi}
\int_0^{2\pi}\frac{d\s}{r^2(\s)+1}\right)\right].\ea

In the particular case of {\it conformal gauge}, (\ref{ae2}) -
(\ref{ss2}) reproduce the corresponding results derived in
\cite{17}. To write down the solutions for $X^\mu$ as functions of
$\tau$ and $\s$ (instead of $\tau$ and $r$), as given in
\cite{17}, it is enough for one to replace (\ref{isr}) into
(\ref{aX}).

In the {\it tensionless limit} $T\to 0$, after using the
$\lambda$-parameterization (\ref{tl}) of $\gamma^{mn}$, one
obtains the following exact string solutions \ba\nl &&\s(r)_{T=0}=
\frac{\left(\lambda^1R\right)^2}{4\lambda^0}
\int\frac{dr^2}{\sqrt{\left(
\mathcal{P}_0^2-\mathcal{P}_2^2\right)r^2 - \mathcal{P}_2^2}}
=\frac{\left(\lambda^1R\right)^2}{2\lambda^0 \left(
\mathcal{P}_0^2-\mathcal{P}_2^2\right)} \sqrt{\left(
\mathcal{P}_0^2-\mathcal{P}_2^2\right)r^2 - \mathcal{P}_2^2},\\
\nl &&r^2(\s)_{T=0}= \frac{\left(2\lambda^0\right)^2\left(
\mathcal{P}_0^2-\mathcal{P}_2^2\right)}{\left(\lambda^1R\right)^4}
\sigma^2 +
\frac{\mathcal{P}_2^2}{\mathcal{P}_0^2-\mathcal{P}_2^2};\ea \ba\nl
&&X^0(\tau,r)_{T=0}\equiv t(\tau,r)_{T=0}= c_1\left[\tau+
\frac{\s(r)}{\lambda^1}\right]- \frac{\mathcal{P}_0}{2}
\int\frac{dr^2}{(r^2+1)\sqrt{\left(
\mathcal{P}_0^2-\mathcal{P}_2^2\right)r^2 - \mathcal{P}_2^2}},\\
\nl &&X^2(\tau,r)_{T=0}\equiv \phi(\tau,r)_{T=0}=
-c_1\frac{\mathcal{P}_0}{\mathcal{P}_2}\left[\tau+
\frac{\s(r)}{\lambda^1}\right]+ \frac{\mathcal{P}_2}{2}
\int\frac{dr^2}{r^2\sqrt{\left(
\mathcal{P}_0^2-\mathcal{P}_2^2\right)r^2 - \mathcal{P}_2^2}};\ea
\ba\nl &&X^0(\tau,\s)_{T=0}\equiv t(\tau,\s)_{T=0}=
c_1\left(\tau+\s/\lambda^1\right) -
\frac{2\lambda^0\mathcal{P}_0}{\left(\lambda^1R\right)^2}
\int\frac{d\s}{r^2(\s)+1},\\ \nl &&X^2(\tau,\s)_{T=0}\equiv
\phi(\tau,\s)_{T=0}=
-c_1\frac{\mathcal{P}_0}{\mathcal{P}_2}\left(\tau+\s/\lambda^1\right)
+ \frac{2\lambda^0\mathcal{P}_2}{\left(\lambda^1R\right)^2}
\int\frac{d\s}{r^2(\s)}.\ea

\subsection{Obtaining new solutions}

Let us give an example, which is not discussed in the
literature yet. To compare with what is known, let us recall that
in \cite{8}, the string dynamics is considered in the background
of $AdS_5$ black hole, with field theory dual {\it finite
temperature} $\mathcal{N}=4$ $SYM$. The background metric is taken
to be \ba\label{fb} &&ds^2_{AdS_5BH}=
-\left(1+\frac{r^2}{R^2}-\frac{M}{r^2}\right)dt^2 +
\frac{dr^2}{\left(1+\frac{r^2}{R^2}-\frac{M}{r^2}\right)}\\ \nl
&&\h\h\h\h + r^2\left(d\theta^2 + \sin^2\theta d\phi_1^2 +
\cos^2\theta d\phi_2^2 \right).\ea The classical string
configuration, investigated in \cite{8} in this gravitational
field, is given by the ansatz \ba\label{la3} t=\tau,\h r=r(\s),\h
\theta=\frac{\pi}{2},\h \phi_1=\omega\tau,\h \phi_2=0.\ea The
background felt by the string for this embedding is \ba\nl ds^2=
-\left(1+\frac{r^2}{R^2}-\frac{M}{r^2}\right)dt^2 +
\frac{dr^2}{\left(1+\frac{r^2}{R^2}-\frac{M}{r^2}\right)}+ r^2
d\phi_1^2,\ea and it depends on $x^1=r$ only. Therefore,
$\mu=(0,2)$, $a=1$ in our notation. (\ref{la3}) is particular case
of (\ref{sLA}), corresponding to \ba\nl \Lambda_0^0=1,\h
\Lambda_0^2=\omega,\h \Lambda_1^\mu=0,\h Z^1(\s)=r(\s).\ea

Now, we are going to find {\it new} exact string solution, based
on the ansatz (\ref{sGA}), which corresponds to more general
string embedding than (\ref{sLA}). Moreover, none of the
coordinates in (\ref{fb}) will be kept fixed. As far as the string
background (\ref{fb}) depends on $x^1=r$ and $x^2=\theta$,
$\mu=(0,3,4)$, $a=(1,2)$ in our notation. In this case, the ansatz
(\ref{sGA}) reduces to (we take $\Lambda_1^\mu=0$): \ba\nl
&&X^0(\tau,\sigma)\equiv t(\tau,\sigma) = \Lambda_0^0\tau +
Z^0(\sigma),\\ \nl &&X^1(\tau,\sigma)= Z^1(\sigma)=r(\sigma), \\
\label{sGA2} &&X^2(\tau,\sigma)= Z^2(\sigma)=\theta(\sigma), \\ \nl
&&X^3(\tau,\sigma)\equiv \phi_1(\tau,\sigma) = \Lambda_0^3\tau +
Z^3(\sigma),\\ \nl &&X^4(\tau,\sigma)\equiv \phi_2(\tau,\sigma) =
\Lambda_0^4\tau + Z^4(\sigma).\ea

The explicit expressions for the conserved momenta $P_\mu$ can be
found from (\ref{mgcm}). By using them, one receives: \ba\nl
&&E=-\int_{0}^{2\pi}d\s P_0(\s) = -T\sqrt{-\gamma}
\int_{0}^{2\pi}d\s\left(1+\frac{r^2}{R^2}-\frac{M}{r^2}\right)
\left(\gamma^{00}\Lambda_0^0 + \gamma^{01}Z'^0\right),
\\ \label{cs31} &&S_1=\int_{0}^{2\pi}d\s P_3(\s) = -
T\sqrt{-\gamma}\int_{0}^{2\pi}d\s r^2\sin^2\theta
\left(\gamma^{00}\Lambda_0^3 + \gamma^{01}Z'^3\right),\\ \nl
&&S_2=\int_{0}^{2\pi}d\s P_4(\s) = -
T\sqrt{-\gamma}\int_{0}^{2\pi}d\s r^2\cos^2\theta
\left(\gamma^{00}\Lambda_0^4 + \gamma^{01}Z'^4\right).\ea

The first integrals $\mathcal{P}_\mu$ for this part of the
equations of motion (\ref{mgaem}), which  correspond to $L=\mu$,
are given by (\ref{mgcq}). Now, they simplify to: \ba\nl
\mathcal{P}_\mu = - T\sqrt{-\gamma}g_{\mu\nu}
\left(\gamma^{11}Z'^\nu + \gamma^{01}\Lambda_0^\nu \right)
=constants.\ea The connection between the conserved momenta
$P_\mu$ and the constants of the motion $\mathcal{P}_\mu$
following from (\ref{cmcq}), allows one to exclude $Z'^\mu$ from
(\ref{cs31}) and to rewrite $E$, $S_1$ and $S_2$ as follows \ba\nl
&&E= -2\pi\frac{\gamma^{01}}{\gamma^{11}}\mathcal{P}_0
+\frac{T\Lambda_0^0}{\sqrt{-\gamma}}
\int_{0}^{2\pi}d\s\left(1+\frac{r^2}{R^2}-\frac{M}{r^2}\right),
\\ \nl &&S_1= 2\pi\frac{\gamma^{01}}{\gamma^{11}}\mathcal{P}_3
+\frac{T\Lambda_0^3}{\sqrt{-\gamma}} \int_{0}^{2\pi}d\s
r^2\sin^2\theta ,\\ \nl &&S_2=
2\pi\frac{\gamma^{01}}{\gamma^{11}}\mathcal{P}_4
+\frac{T\Lambda_0^4}{\sqrt{-\gamma}} \int_{0}^{2\pi}d\s
r^2\cos^2\theta .\ea It is easy to check, that the above three
quantities are constrained by the equality: \ba\nl &&E=
-2\pi\frac{\gamma^{01}}{\gamma^{11}}\left[ \mathcal{P}_0 +
\frac{\Lambda_0^0}{R^2}\left( \frac{\mathcal{P}_3}{\Lambda_0^3} +
\frac{\mathcal{P}_4}{\Lambda_0^4 }\right)\right] +
\frac{\Lambda_0^0}{R^2}\left( \frac{S_1}{\Lambda_0^3} +
\frac{S_2}{\Lambda_0^4 }\right)\\ \nl &&\h\h+\frac{2\pi
T\Lambda_0^0}{\sqrt{-\gamma}}\left(1-
\frac{M}{2\pi}\int_0^{2\pi}\frac{d\s}{r^2(\s)}\right).\ea

Let us now consider the compatibility condition (\ref{cc}). In the
case at hand, it gives \ba\label{cc3} \Lambda_0^4=
-\frac{1}{\mathcal{P}_4}\left(\Lambda_0^0\mathcal{P}_0 +
\Lambda_0^3\mathcal{P}_3\right).\ea From now on, we will use this
expression for $\Lambda_0^4$, thus ensuring that the constraint
(\ref{mga1}) is identically satisfied.

Our next goal is to solve the equations of motion (\ref{mgemf}) and the
constraint (\ref{mgecf}). As explained in Sec.3, these equations have the
same form as (\ref{fem}) and (\ref{ec}). Thus, for obtaining exact string
solutions in the framework of the ansatz (\ref{sGA}), we can use the
formulas derived in Appendix A, after the replacements
$(g,\Gamma,\mathcal{U},\mathcal{A})$ $\to$ $(h,\Gamma^{\bf{h}},
\mathcal{U}^{\bf{h}},\mathcal{A}^{\bf{h}})$. Since the background metric
(\ref{fb}) is diagonal one and $b_{MN}=0$, \ba\nl h_{ab}=g_{ab},\h
\Gamma^{{\bf h}}_{a,bc}= \Gamma_{a,bc},\h \mathcal{A}^{{\bf h}}_a =0.\ea
So, we only need to replace $\mathcal{U}\to\mathcal{U}^{{\bf h}}$ and
$\mathcal{A}\to\mathcal{A}^{\bf{h}}=0 $ in (\ref{eed}) and (\ref{ecd}). As
far as the diagonal background in (\ref{fb}) depends on two coordinates,
we can use the general expressions (\ref{fia}) and (\ref{fir}) for the
first integrals of the equations (\ref{eeda}), (\ref{eedr}), which also
solve the constraint (\ref{ecd}), if the derived sufficient conditions
(\ref{egfr}), (\ref{ca}) and (\ref{cr}) are satisfied. Let us check this.
The condition (\ref{egfr}) does not appear, because in our case the
effective gauge potential $\mathcal{A}^{\bf{h}}$ is identically zero. The
first of the conditions (\ref{ca}) is satisfied, because it takes the form
\ba\nl \frac{\p}{\p\theta} \left[ \frac{g_{22}(r)}{g_{11}(r)}\right]\equiv
0.\ea Consequently, it remains to satisfy the second of the conditions
(\ref{ca}) and the condition (\ref{cr}). In the case at hand, they
require, the right hand sides of (\ref{fia}) and (\ref{fir}) to depend
only on $Z^2=\theta$ and $Z^1=r$ respectively. To see if this is true, let
us write down the first integrals (\ref{fir}) and (\ref{fia}) explicitly
\ba\label{fir'} &&\left(g_{11}r'\right)^2 =
-\frac{g_{11}(r)}{g_{22}(r)}D_2(r)\equiv F_1(r)\ge 0, \\ \nl
&&\left(g_{22}\theta'\right)^2 = D_2(r) +
g_{22}(r)\mathcal{U}^{\bf{h}}(r,\theta)\\ \nl &&= D_2(r) -
\frac{1}{\gamma\left(\gamma^{11}\right)^2}\left\{
\frac{\left(\Lambda_0^0\right)^2}{R^2}\Delta(r) -
\left[\left(\Lambda_0^3\right)^2\sin^2\theta +
\frac{\left(\Lambda_0^0\mathcal{P}_0 +
\Lambda_0^3\mathcal{P}_3\right)^2}{\left(\mathcal{P}_4\right)^2}
\cos^2\theta\right]r^4 \right.\\ \nl &&+ \left.\frac{1}{T^2}\left[
\frac{\mathcal{P}_0^2 R^2r^4}{\Delta(r)} -
\left(\frac{\mathcal{P}_3^2}{\sin^2\theta} +
\frac{\mathcal{P}_4^2}{\cos^2\theta}\right)\right]\right\}\ge 0,
\\ \nl &&\Delta(r)= r^4+R^2(r^2-M).\ea It is
evident that the r.h.s. of the equation for $r'$ is a function only
on $r$, while the r.h.s. of the equation for $\theta'$ is not a
function only on $\theta$. However, we have enough freedom to
satisfy this last condition. To this end, we choose the arbitrary
parameter $\Lambda_0^3$ and the arbitrary function $D_2(r)$ to be
given by \ba\nl &&\Lambda_0^3 =
-\frac{\mathcal{P}_0}{\mathcal{P}_3\pm\mathcal{P}_4}\Lambda_0^0 \h
\Rightarrow \h \Lambda_0^4 = -\frac{\mathcal{P}_0}{\mathcal{P}_4}
\left(1 - \frac{1}{\mathcal{P}_3\pm\mathcal{P}_4}\right)
\Lambda_0^0,
\\ \label{d2r} &&D_2(r) = c^2 +
\frac{1}{\gamma\left(\gamma^{11}\right)^2}\left[
\frac{\left(\Lambda_0^0\right)^2}{R^2}\Delta(r) -
\frac{\left(\Lambda_0^0\mathcal{P}_0\right)^2} {\left(\mathcal{P}_3
\pm\mathcal{P}_4\right)^2}r^4 + \frac{1}{T^2}\frac{\mathcal{P}_0^2
R^2r^4}{\Delta(r)}\right]\le 0,\ea where $c$ is arbitrary constant. After
this choice, the first integral of the equation of motion for the string
coordinate $\theta(\s)$ takes the form \ba\label{fit'}
\left(g_{22}\theta'\right)^2 = c^2 +
\frac{1}{\gamma\left(\gamma^{11}\right)^2T^2}
\left(\frac{\mathcal{P}_3^2}{\sin^2\theta} +
\frac{\mathcal{P}_4^2}{\cos^2\theta}\right)\equiv F_2(\theta)\ge 0.\ea In
this way, all conditions are satisfied, and we can start solving the
equations (\ref{fir'}) and (\ref{fit'}). The general solution of
(\ref{fir'}) is given by \ba\nl \s(r)= R^2 \int\frac{r^2 d r}{\Delta(r)
\sqrt{F_1(r)}}.\ea If we can obtain $r(\s)$ from here, the general
solution of (\ref{fit'}) will be \ba\nl
\int\frac{d\theta}{\sqrt{F_2(\theta)}} = \int\frac{d\s}{r^2(\s)}.\ea
Anyway, we can always find the orbit $r=r(\theta)$ from (\ref{fir'}) and
(\ref{fit'}), and it is defined by the equality \ba\label{orb} R^2
\int\frac{d r}{\Delta(r) \sqrt{F_1(r)}} =
\int\frac{d\theta}{\sqrt{F_2(\theta)}}.\ea

To find the general solution for the remaining string coordinates
$X^\mu$, we have to integrate the equations (\ref{muc}), which in
our case simplify to \ba\label{zmu} Z'^\mu =
-\frac{\gamma^{01}}{\gamma^{11}}\Lambda^\mu_0 -
\frac{\mathcal{P}_\nu}{T\sqrt{-\gamma}\gamma^{11}}
\left(g^{-1}\right)^{\mu\nu},\ea and then replace the received
$Z^\mu$ into the initial ansatz (\ref{sGA2}). The same result may
be obtained directly from (\ref{aX}) for solutions of the type
$X^\mu(\tau,\s)$. Written explicitly, they are: \ba\nl
&&X^0(\tau,\sigma)\equiv t(\tau,\sigma) = \Lambda_0^0\left(\tau -
\frac{\gamma^{01}}{\gamma^{11}}\s\right) +
\frac{R^2\mathcal{P}_0}{T\sqrt{-\gamma}\gamma^{11}}
\int\frac{r^2(\s)d\s}{\Delta[r(\s)]},\\ \nl
&&X^3(\tau,\sigma)\equiv \phi_1(\tau,\sigma) = -
\frac{\Lambda_0^0\mathcal{P}_0}{\mathcal{P}_3\pm\mathcal{P}_4}
\left(\tau - \frac{\gamma^{01}}{\gamma^{11}}\s\right) -
\frac{\mathcal{P}_3}{T\sqrt{-\gamma}\gamma^{11}}
\int\frac{d\s}{r^2(\s)\sin^2\theta(\s)},\\ \nl
&&X^4(\tau,\sigma)\equiv \phi_2(\tau,\sigma) = -
\frac{\Lambda_0^0\mathcal{P}_0}{\mathcal{P}_4} \left(1-
\frac{1}{\mathcal{P}_3\pm\mathcal{P}_4}\right)\left(\tau -
\frac{\gamma^{01}}{\gamma^{11}}\s\right)\\ \nl &&\hspace{4cm}-
\frac{\mathcal{P}_4}{T\sqrt{-\gamma}\gamma^{11}}
\int\frac{d\s}{r^2(\s)\cos^2\theta(\s)}.\ea It is easy to take
{\it conformal gauge} or the {\it tensionless limit} $T\to 0$ in
the above solutions, and we will not consider these cases
separately here.

As a final example, let us obtain a family of {\it new} string
solutions in $AdS_5\times S^5$ background with two spins and up to
{\it nine} independent conserved $R$-charges. We take the line
elements in $AdS_5$ and on $S^5$ to be given by \ba\nl
&&ds^2_{AdS_5}=R^2\left[-\cosh^2\rho dt^2+d\rho^2+\sinh^2\rho
\left(d\theta^2+\sin^2\theta d\phi^2+\cos^2\theta
d\varphi^2\right)\right],\\ \nl
&&ds^2_{S^5}=R^2\left[d\gamma^2+\cos^2\gamma d\varphi^2_3
+\sin^2\gamma\left(d\psi^2+\cos^2\psi d\varphi^2_1 +\sin^2\psi
d\varphi^2_2\right)\right].\ea Now, consider the following string
embedding of type (\ref{sLA}) (the angular coordinates $\theta$
and $\gamma$ are kept fixed to $\pi/4$) \ba\nl
&&X^0(\tau,\sigma)\equiv t(\tau,\sigma) = \Lambda_0^0\tau +
\Lambda^0_1\sigma,\h X^4(\tau,\sigma)=Z^4(\s)=\psi(\s),
\\ \nl &&X^1(\tau,\sigma)=Z^1(\sigma)=\rho(\sigma),\h\h\h\hspace{.2cm}
X^5(\tau,\sigma)\equiv \varphi_1(\tau,\s)=
\Lambda_0^5\tau+\Lambda_1^5\s,
\\ \label{sLA2} &&X^2(\tau,\sigma)\equiv \phi(\tau,\sigma)=
\Lambda_0^2\tau + \Lambda_1^2\sigma,\h X^6(\tau,\sigma)\equiv
\varphi_2(\tau,\s)= \Lambda_0^6\tau+\Lambda_1^6\s, \\ \nl
&&X^3(\tau,\sigma)\equiv \varphi(\tau,\sigma)= \Lambda_0^3\tau +
\Lambda_1^3\sigma,\h X^7(\tau,\sigma)\equiv \varphi_3(\tau,\s)=
\Lambda_0^7\tau+\Lambda_1^7\s, \\ \nl &&\mu=(0,2,3,5,6,7),\h
a=(1,4).\ea The background metric seen by the string is \ba\nl
ds^2&=&R^2\left[-\cosh^2\rho dt^2+d\rho^2+\frac{1}{2}\sinh^2\rho
\left(d\phi^2+d\varphi^2\right)\right.\\ \nl &+& \left.\frac{1}{2}
\left(d\psi^2+\cos^2\psi d\varphi^2_1 +\sin^2\psi d\varphi^2_2+
d\varphi^2_3\right)\right].\ea It does not depend on $x^0=t$,
$x^2=\phi$, $x^3=\varphi$, $x^5=\varphi_1$, $x^6=\varphi_2$,
$x^7=\varphi_3$, so the Lagrangian density (\ref{LRa}) does not
depend on $X^\mu(\tau,\s)$, $\mu=(0,2,3,5,6,7)$. Consequently, the
corresponding generalized momenta (\ref{cmla}) are conserved
\ba\nl p_\mu = \int_0^{2\pi}d\s P_\mu(\s)=constants,\ea \ba\nl
&&p_0\equiv -E= R^2T \sqrt{- \gamma}\gamma ^{0n}\Lambda_n^0
\int_0^{2\pi}d\s\cosh^2\rho(\s),\ea\ba \nl &&p_2\equiv S_1=
-\frac{R^2T}{2} \sqrt{- \gamma}\gamma ^{0n}\Lambda_n^2
\int_0^{2\pi}d\s\sinh^2\rho(\s),\\ \nl &&p_3\equiv S_2= -
\frac{R^2T}{2}\sqrt{- \gamma}\gamma ^{0n}\Lambda_n^3
\int_0^{2\pi}d\s\sinh^2\rho(\s), \ea\ba \nl &&p_5\equiv J_1=
-\frac{R^2T}{2}\sqrt{- \gamma}\gamma ^{0n} \Lambda_n^5
\int_0^{2\pi}d\s\cos^2\psi(\s),\\ \label{cmle} &&p_6\equiv J_2=
-\frac{R^2T}{2} \sqrt{- \gamma}\gamma
^{0n}\Lambda_n^6\int_0^{2\pi}d\s \sin^2\psi(\s),\\ \nl &&p_7\equiv
J_3=-\pi R^2T\sqrt{- \gamma}\gamma ^{0n}\Lambda_n^7.\ea In
addition, there exist the constants of the motion (\ref{cq}),
which must satisfy the condition (\ref{cc}). In our case, we may
choose for instance \ba\nl
\Lambda_0^3=-\frac{1}{\mathcal{P}_3}\left(\Lambda_0^0\mathcal{P}_0
+ \Lambda_0^2\mathcal{P}_2 + \Lambda_0^5\mathcal{P}_5 +
\Lambda_0^6\mathcal{P}_6 + \Lambda_0^7\mathcal{P}_7 \right),\ea to
ensure that the constraint (\ref{a1}) is identically fulfilled.

The first integrals (\ref{fir}), (\ref{fia}), of the equations of
motion for $\rho(\s)$ and $\psi(\s)$, which also solve the
constraint (\ref{ecd}), read \ba\label{firl}
&&\left(g_{11}\rho'\right)^2 = -2D_4(\rho)\equiv F_1(\rho)\ge 0,
\\ \nl &&\left(g_{44}\psi'\right)^2 = D_4(\rho) +
g_{44}\mathcal{U}(\rho,\psi)\\ \label{fial} &&= d +
\frac{R^4}{4\gamma^{11}}\left(C^{55}\cos^2\psi + C^{66}\sin^2\psi
+ C^{77}\right)\equiv F_4(\psi)\ge 0,\\ \nl && d=const,\h
C^{\mu\nu}=\gamma^{mn}\Lambda^\mu_m\Lambda^\nu_n.\ea Here, the
arbitrary function $D_4(\rho)$ has been chosen to be given by
\ba\nl D_4(\rho)=d - \frac{R^4}{2\gamma^{11}}
\left[\frac{2\Lambda_1^\mu\mathcal{P}_\mu}{R^2T\sqrt{-\gamma}}
-C^{00}\cosh^2\rho +\frac{1}{2}
\left(C^{22}+C^{33}\right)\sinh^2\rho \right]\le 0,\ea in order
the integrability condition (\ref{cr}) to be satisfied
\footnote{The condition (\ref{egfr}) does not appear, because
$\mathcal{A}_a=0$. The remaining conditions (\ref{ca}) are
satisfied identically.}. The general solutions of the differential
equations (\ref{firl}) and (\ref{fial}), for the string
coordinates $\rho(\s)$ and $\psi(\s)$, are \ba\nl
\s(\rho)=R^2\int\frac{d\rho}{\sqrt{F_1(\rho)}},\h
\s(\psi)=\frac{R^2}{2}\int\frac{d\psi}{\sqrt{F_4(\psi)}}.\ea From
(\ref{firl}) and (\ref{fial}), one can also find the orbit
$\rho=\rho(\psi)$ \ba\nl 2\int\frac{d\rho}{\sqrt{F_1(\rho)}}=
\int\frac{d\psi}{\sqrt{F_4(\psi)}}.\ea

Now, let us show that for the string solution obtained above,
there exist other nontrivial conserved angular momenta on $S^5$
($R$-charges), besides $p_5=J_1$, $p_6=J_2$ and $p_7=J_3$. To this
aim, following \cite{30}, we introduce new embedding coordinates
on $S^5$, working in {\it conformal gauge} \ba\nl &&W_1+i
W_2=R\sin\gamma\cos\psi e^{i\varphi_1},\h W_3+i W_4
=R\sin\gamma\sin\psi e^{i\varphi_2},\\ \nl &&W_5+i W_6
=R\cos\gamma e^{i\varphi_3},\h\sum_{A=1}^{6}W_A^2=R^2.\ea In terms
of these coordinates, the generators of the $S^5$ isometry group
$O(6)$ are ($\p_0=\p/\p\tau$) \ba\nl
J_{AB}=T\int_{0}^{2\pi}d\s\left(W_A\p_0 W_B-W_B\p_0 W_A\right),\ea
and they must be conserved in time: $\p_0 J_{AB}=0$.

For our ansatz (\ref{sLA2}), the coordinates $W_A$ acquire the
form \ba\nl &&W_1(\tau,\s)=\frac{R}{\sqrt{2}}\cos\psi(\s)
\cos\left(\Lambda_0^5\tau+\Lambda_1^5\s\right),\hspace{.2cm}
W_2(\tau,\s)=\frac{R}{\sqrt{2}}\cos\psi(\s)
\sin\left(\Lambda_0^5\tau+\Lambda_1^5\s\right),\\ \nl
&&W_3(\tau,\s)=\frac{R}{\sqrt{2}}\sin\psi(\s)
\cos\left(\Lambda_0^6\tau+\Lambda_1^6\s\right),\hspace{.2cm}
W_4(\tau,\s)=\frac{R}{\sqrt{2}}\sin\psi(\s)
\sin\left(\Lambda_0^6\tau+\Lambda_1^6\s\right),\\ \nl
&&W_5(\tau,\s)=\frac{R}{\sqrt{2}}
\cos\left(\Lambda_0^7\tau+\Lambda_1^7\s\right),\hspace{.2cm}
W_6(\tau,\s)=\frac{R}{\sqrt{2}}
\sin\left(\Lambda_0^7\tau+\Lambda_1^7\s\right).\ea By using these
expressions for $W_A$, it is easy to check that $J_{12}$, $J_{34}$
and $J_{56}$ coincide with the conserved momenta (\ref{cmle}),
taken in conformal gauge \ba\nl &&J_{12}=p_5^{cg}=
\frac{R^2T}{2}\Lambda_0^5 \int_0^{2\pi}d\s\cos^2\psi(\s),\\ \nl
&&J_{34}=p_6^{cg}= \frac{R^2T}{2}\Lambda_0^6
\int_0^{2\pi}d\s\sin^2\psi(\s),\\ \nl &&J_{56}=p_7^{cg}=\pi
R^2T\Lambda_0^7.\ea The conservation conditions $\p_0 J_{AB}=0$,
applied to the remaining angular momenta $J_{AB}$, constrain the
parameters $\Lambda_0^5$, $\Lambda_0^6$ and $\Lambda_0^7$ as
follows \ba\nl \Lambda_0^5=\Lambda_0^7=\omega,\h
\Lambda_0^6=\pm\omega.\ea Taking this into account, one obtains
the following expressions for the other generators $J_{AB}$ \ba\nl
J_{13}^{\pm}&=&\pm
J_{24}^{\pm}=\frac{R^2T}{2}\omega\int_0^{2\pi}d\s
\left[\sin\left(\omega\tau+\Lambda_1^5\s\right)
\cos\left(\pm\omega\tau+\Lambda_1^6\s\right)\right.\\ \nl &\mp&
\left.\sin\left(\pm\omega\tau+\Lambda_1^6\s\right)
\cos\left(\omega\tau+\Lambda_1^5\s\right)\right]
\sin\psi(\s)\cos\psi(\s)\\ \nl J_{14}^{\pm}&=&\mp
J_{23}^{\pm}=\frac{R^2T}{2}\omega\int_0^{2\pi}d\s
\left[\sin\left(\omega\tau+\Lambda_1^5\s\right)
\sin\left(\pm\omega\tau+\Lambda_1^6\s\right)\right.\\ \nl &\pm&
\left.\cos\left(\omega\tau+\Lambda_1^5\s\right)
\cos\left(\pm\omega\tau+\Lambda_1^6\s\right)\right]
\sin\psi(\s)\cos\psi(\s)\\ \nl J_{15}&=&
J_{26}=\frac{R^2T}{2}\omega\int_0^{2\pi}d\s
\left[\sin\left(\omega\tau+\Lambda_1^5\s\right)
\cos\left(\omega\tau+\Lambda_1^7\s\right)\right.\\ \nl &-&
\left.\sin\left(\omega\tau+\Lambda_1^7\s\right)
\cos\left(\omega\tau+\Lambda_1^5\s\right)\right]\cos\psi(\s)\\ \nl
J_{16}&=&-J_{25}=\frac{R^2T}{2}\omega\int_0^{2\pi}d\s
\left[\sin\left(\omega\tau+\Lambda_1^5\s\right)
\sin\left(\omega\tau+\Lambda_1^7\s\right)\right.\\ \nl &+&
\left.\cos\left(\omega\tau+\Lambda_1^5\s\right)
\cos\left(\omega\tau+\Lambda_1^7\s\right)\right]\cos\psi(\s)\\ \nl
J_{35}^{\pm}&=&\pm
J_{46}^{\pm}=\pm\frac{R^2T}{2}\omega\int_0^{2\pi}d\s
\left[\sin\left(\pm\omega\tau+\Lambda_1^6\s\right)
\cos\left(\omega\tau+\Lambda_1^7\s\right)\right.\\ \nl &\mp&
\left.\sin\left(\omega\tau+\Lambda_1^7\s\right)
\cos\left(\pm\omega\tau+\Lambda_1^6\s\right)\right]\sin\psi(\s)\\
\nl  J_{36}^{\pm}&=&\mp
J_{45}^{\pm}=\pm\frac{R^2T}{2}\omega\int_0^{2\pi}d\s
\left[\sin\left(\pm\omega\tau+\Lambda_1^6\s\right)
\sin\left(\omega\tau+\Lambda_1^7\s\right)\right.\\ \nl &\pm&
\left.\cos\left(\pm\omega\tau+\Lambda_1^6\s\right)
\cos\left(\omega\tau+\Lambda_1^7\s\right)\right]\sin\psi(\s).\ea
Hence, our string solution characterizes with {\it nine
independent conserved} $R$-charges, if $\Lambda_1^5$,
$\Lambda_1^6$, $\Lambda_1^7$, are all different.

\setcounter{equation}{0}
\section{Concluding remarks}
The string embeddings (\ref{tLA}) - (\ref{sGA}) allow us to reduce
the problem of solving the string equations of motion and
constraints to a particle-like one. To achieve this, one have to
get rid of the dependence on the spatial worldsheet coordinate
$\s$, or on the temporal worldsheet coordinate $\tau$. To this
aim, since the string action contains the first derivatives $\p_m
X^M$, the string coordinates $X^M(\tau,\s)$ must depend on $\s$ or
$\tau$ at most linearly. Besides, the background fields entering
the action depend implicitly on $\tau$ and $\s$ through their
dependence on $X^M$. If we accept that the external fields depend
only on part of the coordinates, say $X^a$, the resulting reduced
Lagrangian density will depend only on $\xi^0=\tau$ for the
ansatzes (\ref{tLA}) and (\ref{tGA}), or on $\xi^1=\s$ for the
ansatzes (\ref{sLA}) and (\ref{sGA}).

Of course, one can use different approach to simplify the problem
of solving the nonlinear string equations of motion in variable
external fields \cite{21,34}. The two ansatzes used in \cite{21},
which do not belong to the types (\ref{tLA}) - (\ref{sGA}), are
the following \ba\nl t=t(\tau),\h r=f(\tau)g(\s),\h
l=F(\tau)G(\s),\h \phi=\omega\tau ,\ea and \ba\nl t=t(\tau),\h
r=r(\tau),\h \theta=\theta(\s),\h \varphi=\nu\tau.\ea The
backgrounds do not depend on $(t,\phi)$ and on $(t,\varphi)$
respectively. The three ansatzes, used in \cite{34}, are \ba\nl
&&t=c_0\tau,\h r=r(\s),\h \phi=\omega\tau,\h
\theta=\theta(\tau),\h \psi=\sigma,\\ \nl &&t=t(\tau),\h
r=r(\tau),\h \phi=\sigma,\h \theta=\theta(\sigma),\h
\psi=\nu\tau,\\ \nl &&t=c_0\tau,\h \beta=\sigma,\h
\phi=\omega\tau,\h \tilde{\phi}=\omega\tau,\h
\theta=\theta(\tau),\h \psi=\sigma .\ea The corresponding
background metrics seen by the string do not depend on
$(t,\phi,\psi)$ in the first two cases and on
$(t,\phi,\tilde{\phi},\psi)$ in the third one. Therefore, all the
ansatzes used in the literature, are particular cases of the
following string embedding \ba\nl X^\mu(\tau,\sigma)=\Lambda_0^\mu
\tau+\Lambda_1^\mu \sigma + Y^\mu(\tau)Z^\mu(\s),\h
X^a(\tau,\sigma)=Y^a(\tau)Z^a(\s).\ea Let us also mention that the
ansatz of the type $X(\tau,\sigma)=Y(\tau)Z(\sigma)$, used in
\cite{30}, is equivalent to (\ref{sLA}) with $\Lambda_1^\mu=0$,
after change of the embedding coordinates \footnote{See the last
example in the previous section.}.

Our task in this paper was to find as more general string
solutions as we can. That is why, we do not consider the different
periodicity and normalization conditions, which arise for
particular string configurations. However, once the general
solution is found, they can be always implemented.

\vspace*{.5cm} {\bf Acknowledgments} \vspace*{.2cm}

This work is supported by Shoumen University grants under
contracts {\it No.001/2003} and {\it No.003/2004}.


\setcounter{section}{1}
\setcounter{subsection}{0}
\appendix{}

Here, we will explain how to find solutions of the string
equations of motion (\ref{fem}) and of the effective constraint
(\ref{ec}).

Let us start with the simplest case, when the background fields
depend on only one coordinate $X^a(\tau,\sigma)=Z^a(\sigma)$. In
this case the Eqs. (\ref{fem}), (\ref{ec}) simplify to
\ba\label{ee1} &&\frac{d}{d\sigma}(g_{aa}Z'^a)
-\frac{1}{2}\frac{dg_{aa}}{dZ^a}\left(Z'^a\right)^2 =
\frac{1}{2}\frac{d\mathcal{U}}{dZ^a} ,\\ \label{ec1}
&&g_{aa}\left(Z'^a\right)^2 = \mathcal{U},\ea where we have used
that \ba\nl g_{ab}Z''^b + \Gamma_{a,bc}Z'^bZ'^c
=\frac{d}{d\sigma}\left(g_{ab}Z'^b\right) -\frac{1}{2}\p_a
g_{bc}Z'^bZ'^c.\ea After multiplying with $2g_{aa}Z'^a$ and after
using the constraint (\ref{ec1}),  the Eq. (\ref{ee1}) reduces to
\ba\label{fi1} \frac{d}{d\sigma}\left[\left(g_{aa}Z'^a\right)^2 -
g_{aa}\mathcal{U}\right]=0.\ea The solution of (\ref{fi1}),
compatible with (\ref{ec1}), is just the constraint (\ref{ec1}).
In other words, (\ref{ec1}) is first integral of the equation for
the coordinate $Z^a$. By integrating (\ref{ec1}), one obtains the
following exact string solution
\ba\label{tsol1}\sigma\left(X^a\right)=\sigma_0 +
\int_{X_0^a}^{X^a}
\left(\frac{\mathcal{U}}{g_{aa}}\right)^{-1/2}dx,\ea where
$\sigma_0$ and $X_0^a$ are arbitrary constants.

Let us turn to the more complicated case, when the background
fields depend on more than one coordinate
$X^a(\tau,\sigma)=Z^a(\s)$. We would like to apply the same
procedure for solving the system of differential equations
(\ref{fem}), (\ref{ec}), as in the simplest case just considered.
To be able to do this, we need to suppose that the metric $g_{ab}$
is a diagonal one. Then one can rewrite the effective equations of
motion (\ref{fem}) and the effective constraint (\ref{ec}) in the
form \ba\label{eed} &&\frac{d}{d\sigma}\left(g_{aa}Z'^a\right)^2 -
Z'^a\p_a\left(g_{aa}\mathcal{U}\right)
\\ \nl &&+ Z'^a\sum_{b\ne a}
\left[\p_a\left(\frac{g_{aa}}{g_{bb}}\right)
\left(g_{bb}Z'^b\right)^2 -
4\p_{[a}\mathcal{A}_{b]}g_{aa}Z'^b\right] = 0,
\\ \label{ecd}
&&g_{aa}\left(Z'^a\right)^2 +\sum_{b\ne a}
g_{bb}\left(Z'^b\right)^2 = \mathcal{U}.\ea

To find solutions of the above equations without choosing
particular background, we can fix all coordinates $X^a$ except
one. Then the exact string solution is given again by the same
expression (\ref{tsol1}) for $\sigma\left(X^a\right)$.

To find solutions depending on more than one coordinate, we must
impose further conditions on the background fields. Let us show,
how a number of {\it sufficient} conditions, which allow us to
reduce the order of the equations of motion by one, can be
obtained.

First of all, we split the index $a$ in such a way that $Z^r$ is
one of the coordinates $Z^a$, and $Z^{\alpha}$ are the others.
Then we assume that the effective 1-form gauge field
$\mathcal{A}_a$ can be represented in the form  \ba\label{egfr}
\mathcal{A}_a \equiv(\mathcal{A}_r,\mathcal{A}_\alpha)=
(\mathcal{A}_r,\p_\alpha f),\ea i.e., it is oriented along the
coordinate $Z^r$, and the remaining components
$\mathcal{A}_\alpha$ are pure gauges. Now, the Eqs. (\ref{eed})
read \ba\label{eeda} &&\frac{d}{d\sigma}\left(g_{\alpha\alpha}
Z'^\alpha\right)^2 -
Z'^\alpha\p_\alpha\left(g_{\alpha\alpha}\mathcal{U}\right)
\\ \nl &&+Z'^\alpha\left[\p_\alpha\left(\frac{g_{\alpha\alpha}}
{g_{rr}}\right)\left(g_{rr}Z'^r\right)^2
-2g_{\alpha\alpha}\p_\alpha\left(\mathcal{A}_r-\p_r f\right)
Z'^r\right]
\\ \nl &&+ Z'^\alpha\sum_{\beta\ne \alpha}
\p_\alpha\left(\frac{g_{\alpha\alpha}}{g_{\beta\beta}}
\right)\left(g_{\beta\beta}Z'^\beta\right)^2 = 0,
\\ \label{eedr} &&\frac{d}{d\sigma}\left(g_{rr}Z'^r\right)^2 -
Z'^r\p_r\left(g_{rr}\mathcal{U}\right)
\\ \nl &&+ Z'^r\sum_{\alpha}
\left[\p_r\left(\frac{g_{rr}}{g_{\alpha\alpha}}\right)
\left(g_{\alpha\alpha}Z'^\alpha\right)^2
+2g_{rr}\p_{\alpha}\left(\mathcal{A}_r-\p_r f\right)
Z'^\alpha\right] = 0.\ea After imposing the conditions
\ba\label{ca} \p_\alpha\left(\frac{g_{\alpha\alpha}}
{g_{aa}}\right)=0, \h\p_\alpha\left(g_{rr}Z'^r\right)^2=0,\ea the
Eqs. (\ref{eeda}) reduce to \ba\nl
\frac{d}{d\sigma}\left(g_{\alpha\alpha}Z'^\alpha\right)^2
-Z'^\alpha\p_\alpha\left\{g_{\alpha\alpha}\left[\mathcal{U}
+2\left(\mathcal{A}_r-\p_r f\right)Z'^r\right]\right\}=0,\ea which
are solved by \ba\label{fia}
\left(g_{\alpha\alpha}Z'^\alpha\right)^2 =D_{\alpha}
\left(Z^{a\ne\alpha}\right) + g_{\alpha\alpha}\left[\mathcal{U}
+2\left(\mathcal{A}_r-\p_r f\right)Z'^r\right]= F_{\alpha}
\left(Z^{\beta}\right)\ge 0,\ea where $D_{\alpha}$, $F_{\alpha}$
are arbitrary functions of their arguments.
\footnote{$F_{\alpha}=F_{\alpha} \left(Z^{\beta}\right)$ follows
from (\ref{cr}).}

To integrate the Eq. (\ref{eedr}), we impose the condition
\ba\label{cr} \p_r\left(g_{\alpha\alpha}Z'^\alpha\right)^2=0. \ea
After using the second of the conditions (\ref{ca}), the condition
(\ref{cr}), and the already obtained solution (\ref{fia}), the Eq.
(\ref{eedr}) can be recast in the form \ba\nl
&&\frac{d}{d\sigma}\left[\left(g_{rr}Z'^r\right)^2
+2g_{rr}\left(\mathcal{A}_r-\p_r f\right)Z'^r\right]\\ \nl
&&=Z'^r\p_r\left\{g_{rr}\left[(1-n_\alpha)\left(\mathcal{U}
+2\left(\mathcal{A}_r-\p_r f\right)Z'^r\right)
-\sum_{\alpha}\frac{D_{\alpha} \left(Z^{a\ne\alpha}\right)}
{g_{\alpha\alpha}}\right]\right\},\ea where $n_\alpha$ is the
number of the coordinates $Z^\alpha$. The solution of this
equation, compatible with (\ref{fia}) and with the effective
constraint (\ref{ecd}), is \ba\label{fir}
\left(g_{rr}Z'^r\right)^2 = g_{rr} \left[(1-n_\alpha)\mathcal{U}
-2n_\alpha\left(\mathcal{A}_r-\p_r f\right)Z'^r
-\sum_{\alpha}\frac{D_{\alpha} \left(Z^{a\ne\alpha}\right)}
{g_{\alpha\alpha}}\right]= F_r\left(Z^r\right)\ge 0,\ea where
$F_r$ is again an arbitrary function.

Thus, we succeeded to separate the variables $Z'^a$ and to obtain
the first integrals (\ref{fia}), (\ref{fir}) for the equations of
motion (\ref{eed}), when the conditions (\ref{egfr}), (\ref{ca}),
(\ref{cr}) on the background fields are fulfilled. Further
progress is possible, when working with particular background
configurations, allowing for separation of the variables in
(\ref{fia}) and (\ref{fir}).

Let us finally point out that (\ref{egfr}), (\ref{ca}) and
(\ref{cr}) are just an {\it example} of possible integrability
conditions, which are fulfilled in many cases, but in many others
- they are not.


\end{document}